\documentclass{article}
\usepackage[T1]{fontenc} 
\usepackage[utf8]{inputenc} 
\usepackage{ismir,amsmath,cite,url}
\usepackage{graphicx}
\usepackage{color}
\usepackage{mathptmx}
\usepackage{amssymb}
\usepackage{amsthm}
\usepackage{bm,bbold}

\newtheorem*{prop*}{Proposition}

\newtheorem*{cor*}{Corollary}
\usepackage{microtype}
\usepackage{musicography}

\usepackage{lineno}

\title{STONE: Self-supervised Tonality Estimator}

\multauthor
{Yuexuan Kong$^{1, 2}$ \hspace{1cm} Vincent Lostanlen$^2$ \hspace{1cm} Gabriel Meseguer-Brocal$^1$} { \bfseries{Stella Wong$ $ \hspace{1cm} Mathieu Lagrange$^2$ \hspace{1cm} Romain Hennequin$^1$}\\
 $^1$ Deezer Research, Paris, France\\
$^2$ Nantes Université, École Centrale Nantes, CNRS, LS2N, UMR 6004, F-44000 Nantes, France\\
{\tt\small ykong@deezer.com}
}

\def\authorname{Y. Kong, V. Lostanlen, G. Meseguer-Brocal, S. Wong, M. Lagrange and R. Hennequin}

\begin{document}

\maketitle
\begin{abstract}
Although deep neural networks can estimate the key of a musical piece, their supervision incurs a massive annotation effort.
Against this shortcoming, we present STONE, the first self-supervised tonality estimator.
The architecture behind STONE, named ChromaNet, is a convnet with octave equivalence which outputs a ``key signature profile'' (KSP) of 12 structured logits.
First, we train ChromaNet to regress artificial pitch transpositions between any two unlabeled musical excerpts from the same audio track, as measured as cross-power spectral density (CPSD) within the circle of fifths (CoF).
We observe that this self-supervised pretext task leads KSP to correlate with tonal key signature.
Based on this observation, we extend STONE to output a structured KSP of 24 logits, and introduce supervision so as to disambiguate major versus minor keys sharing the same key signature.
Applying different amounts of supervision yields semi-supervised and fully supervised tonality estimators: i.e., Semi-TONEs and Sup-TONEs. 
We evaluate these estimators on FMAK, a new dataset of 5489 real-world musical recordings with expert annotation of 24 major and minor keys.
We find that Semi-TONE matches the classification accuracy of Sup-TONE with reduced supervision and outperforms it with equal supervision.
\end{abstract}
\section{Introduction}
Self-taught musicians can tell whether two pieces go ``in tune'' or ``out of tune''.
To do so, they do not need to know the name of every key \cite{mullensiefen2014musicality}. Meanwhile, in music information retrieval (MIR), current tonality estimators depend on a vocabulary of labels such as $\mathtt{C:maj}$ or $\mathtt{F:min}$.

In this context, we aim to develop models which ``learn by ear'' like humans; i.e., from little or no annotated data.
This goal is justified in practice by the fact that online digital music corpora are larger and more musically diverse than established MIR datasets, yet often lack expert metadata.

To overcome the need for large amount of labeled data, self-supervised learning (SSL) has emerged as an alternative paradigm to supervised learning, with numerous applications in speech and music processing \cite{zhu2021musicbert, spijkervet2021contrastive, desblancs2023zero, meseguer2024}.

The design of pretext tasks is a long-standing issue in SSL for audio.
On one hand, some of them are meant as a pretraining step for general-purpose representation learning: contrastive predictive coding\cite{schneider2019wav2vec}, deep metric learning \cite{saeed2021contrastive}, and self-distillation\cite{niizumi2021byol}, to name a few.
On the other hand, another family of pretext tasks is designed to suit a particular downstream task, such as tempo and pitch estimation\cite{riou2023pesto, quinton2022equivariant}.
In this context, the concept of \emph{equivariance} plays a central role.
Loosely speaking, equivariance means that a certain parametric transformation of the input data forms a simple trajectory in the space of learned representations. Yet, equivariant SSL has never been used to study tonality, for lack of an adequate pretext task. 

The main idea of our paper is that, even so \emph{absolute} key labels are unknown, we can construct paired samples in which \emph{relative} harmonic progressions serve as a learning signal for tonality estimation. Our contributions are:
\begin{description}
    \item [STONE.] To our knowledge, the first SSL framework whose model predictions correlates with key signatures. It comprises a new equivariant neural network named ChromaNet and a noncontrastive loss function based on cross-power spectral density (CPSD).
    \item [Downstream task.] We extend STONE into Semi-TONE, a semi-supervised model that is tailored for 24-way key estimation. Semi-TONE performs on par with a supervised counterpart (Sup-TONE) while reducing dependency on annotated data by 90\%\footnote{Companion website: \url{https://github.com/deezer/stone}}.
    \item [FMAK.] A new large dataset of 5489 real-world music recordings, collected from the Free Music Archive (FMA) and annotated by an expert for 24 major and minor keys, is available for free download\cite{wong2023fmak}\footnote{FMAK was first released at \url{https://zenodo.org/records/12759100} by Wong and Hernandez. We then updated 200 songs' annotations later at \url{https://zenodo.org/records/12759100}, version used in this paper. }.
\end{description}

\section{Related Work}
\label{sec:related-work}

\subsection{Equivariant self-supervised learning in music}
Equivariant SSL learns task-specific embeddings by representing the transformations which underlie its factors of variability: e.g., variations in pitch or tempo \cite{quinton2022equivariant, gfeller2020spice, morais2023tempo}.
In particular, PESTO is a monophonic pitch estimator trained by learning the pitch shift of the same sample\cite{riou2023pesto}. However, its extension to multipitch tracking is an open problem \cite{cwitkowitz2024toward}.

\subsection{Computational models of tonality}
Tonality estimation remains a relatively under-researched field, due to the scarcity of labeled data available for both training and evaluation purposes.
The earliest methods were based on template matching \cite{noland2007signal, pauws2004musical, faraldo2016key, krumhansl2001cognitive}. Later, convolutional neural networks (convnets) and transformers appeared, treating the task as a supervised 24-class classification problem \cite{Wu2020AVA, Wu2022JointCA, Schreiber2019MusicalTA}.
Among studies employing the same dataset, the convnet of Korzeniowski \emph{et al.} \cite{korzeniowski2018genre} achieves the best performance. Against the lack of annotated data, prior work proposed to integrate key estimation with unsupervised autoencoding \cite{Wu2020AVA, Wu2022JointCA}. However, their approach is computationally intensive and its evaluation is limited.

\subsection{Annotated datasets for key estimation}
\label{sub:related-datasets}

Key detection datasets face a conundrum between diversity and reproducibility.
The Million Song Dataset \cite{Bertin-Mahieux2011} offers key labels for diverse commercial music but lacks public audio.
GiantSteps MTG Key (GSMK)\footnote{https://github.com/GiantSteps/giantsteps-mtg-key-dataset}, GiantSteps Key (GSK) \cite{Knees2015TwoDS} and McGill Billboard datasets \cite{burgoyne-2011-expert} offer public data, yet they are restricted in terms of genres and quantity. In particular, GSMK and GSK serve in the training and evaluation of supervised SOTA \cite{korzeniowski2018genre}.

\section{Methods}
\label{sec:pretext}
\begin{figure}
    \centering
    \includegraphics[width=\linewidth]{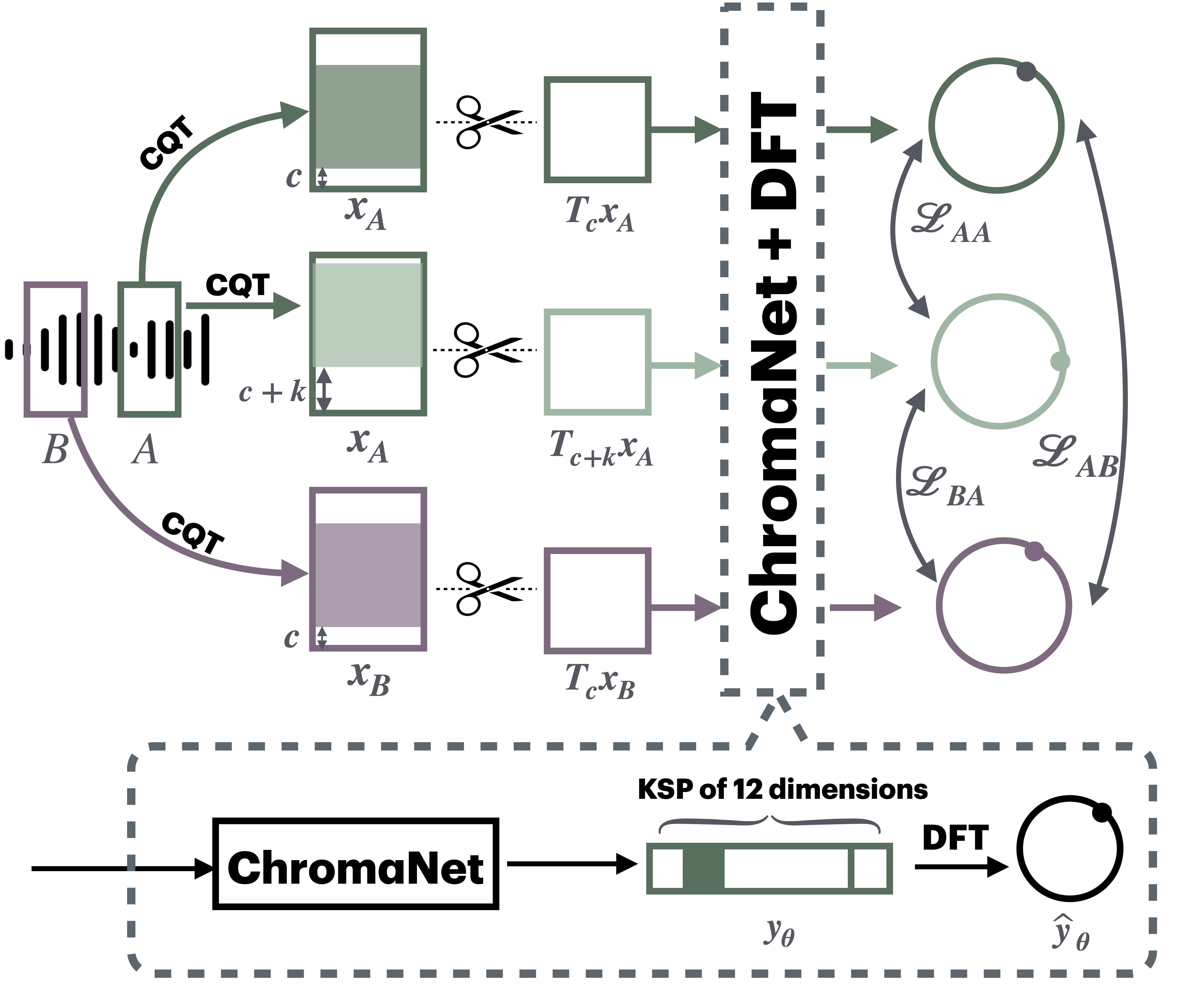}
    \caption{Overview of the equivariant pretext task in STONE. Given two segments A and B from an unlabeled musical recording, we compute their constant-$Q$ transforms (CQT) and apply random crops by $c$ and $(c+k)$ to simulate pitch transpositions. We feed them to ChromaNet, an equivariant neural network with octave equivalence, yielding a learned key signature profile (KSP) of 12 chromas. We compute the discrete Fourier transform (DFT) of each KSP and derive pairwise cross-power spectral densities (CPSD). Self-supervised losses $\mathcal{L}_{\mathrm{AA}}$, $\mathcal{L}_{\mathrm{AB}}$, and $\mathcal{L}_{\mathrm{BA}}$ are formulated as CPSD regression residuals in the complex domain.
    }
    \label{fig:overview}
\end{figure}

Figure \ref{fig:overview} illustrates our proposed method for STONE.

\subsection{Artificial pitch transpositions of the CQT}
\label{sub:cqt}
We compute a constant-$Q$ transform (CQT) with $Q=12$ bins per octave and center frequencies ranging between $\xi_{\min} = 27.5$ Hz and $\xi_{\max} = 2^{99/12} \xi_{\min} = 8.37$ kHz.

Given a CQT matrix $\boldsymbol{x}$ and a integer $c\leq15$, we reduce the number of rows from 99 down to 84 (7 octaves) by trimming the $c$ lowest-frequency bins and $(15-c)$ highest-frequency bins.
This is tantamount by a pitch transposition by $c$ semitones \cite{riou2023pesto}. We denote the result by $T_{c}\boldsymbol{x}$ where $T_{c}\boldsymbol{x}[p, t] = \boldsymbol{x}[p-c, t]$ for each frequency $p<84$ and time $t$.

\subsection{ChromaNet: a convnet with octave equivalence}
\label{sub:chromanet}

The cropped CQT matrix $T_{c} \boldsymbol{x}$ has a frequency range of $QJ=84$ semitones with $Q=12$ and $J=7$ octaves.
We define a 2-D fully convolutional network $f_{\boldsymbol{\theta}}$ with trainable parameters $\boldsymbol{\theta}$, operating on $T_{c}\boldsymbol{x}$ with no pooling over the frequency dimension.
The last layer has a single channel and performs global average pooling over the time dimension.

The architecture $f_{\boldsymbol{\theta}}$ composes seven blocks, each of them composing a ConvNeXT block  \cite{liu2022convnet} and a time downsampling block, and layer normalization.
It returns a vector in dimension $QJ$.
While ConvNeXt blocks leaves the input resolution unchanged, time downsampling blocks decrease time resolution while preserving frequency resolution.

We compose the convnet $f_{\boldsymbol{\theta}}$ with a non-trainable operator $g$ whose role is to guarantee octave equivalence.
We roll the log-frequency axis into a spiral which makes a full turn at every octave, thus aligning coefficients in $f_{\boldsymbol{\theta}}(T_{c}\boldsymbol{x})$ of the form $(p\pm Qj)$ for integer $j$.
The operator $g$ sums these coefficients across octaves $j$ for each pitch class $q$ and applies a softmax transformation.
We obtain a $Q$-dimensional vector $\boldsymbol{y}_{\boldsymbol{\theta}}$ whose entries are nonnegative and sum to one.
We propose to call this vector a ``key signature profile'' (KSP):
\begin{align}
\boldsymbol{y}_{\boldsymbol{\theta}}[q] &=
(g \circ f_{\boldsymbol{\theta}})(T_{c}\boldsymbol{x})[q] \nonumber \\
&=
\dfrac{
\exp\left(\sum_{j=0}^{J-1} f_{\boldsymbol{\theta}}(T_{c}\boldsymbol{x})[Qj+q]\right)
}{
\sum_{q'=0}^{Q-1}\exp\left(\sum_{j=0}^{J-1} f_{\boldsymbol{\theta}}(T_{c}\boldsymbol{x})[Qj+q']\right)
}.
\label{eq:softmax}
\end{align}
For brevity, we do not recall the dependency of $\boldsymbol{y}_{\boldsymbol{\theta}}$ upon $\boldsymbol{x}$ nor $c$.
Equation 1 resembles the extraction of chroma features \cite{muller2011signal}, hence the proposed name of ChromaNet.

\subsection{DFT over key signature profiles}

With $Q=12$, the discrete Fourier transform (DFT) of the KSP $\boldsymbol{y}_{\boldsymbol{\theta}}$ is
\begin{equation}
\widehat{\boldsymbol{y}}_{\boldsymbol{\theta}}[\omega]
= \mathcal{F}\{\boldsymbol{y}_{\boldsymbol{\theta}}\}[\omega] = \sum_{q=0}^{11} \boldsymbol{y}_{\boldsymbol{\theta}}[q] e^{-2\pi\mathrm{i}\omega q/12},
\label{eq:fourier-transform}
\end{equation}
where $\omega$ is an integer between $0$ and $11$ that is coprime to 12 for a full circular distribution over all 12 pitches.
With $\omega = 7$, a circular shift of $\boldsymbol{y}_{\boldsymbol{\theta}}$ by seven chromas corresponds to a multiplication of $\widehat{\boldsymbol{y}}_{\boldsymbol{\theta}}[\omega]$ by $e^{2\pi\mathrm{i}49/12}=e^{2\pi\mathrm{i}/12}$.
Hence, the phase of the complex number $\widehat{\boldsymbol{y}}_{\boldsymbol{\theta}}[\omega]$ denotes a key modulation in the circle of fifths (CoF).
Alternatively, $\omega = 1$ would correspond to a circle of semitones. Our paper evaluates both  settings but only describes the CoF setting ($\omega=7$) for the sake of conciseness. 

\subsection{Cross-power spectral density (CPSD)}
Let us split the CQT matrix $\boldsymbol{x}$ into two disjoint time segments of equal length: $\boldsymbol{x} = (\boldsymbol{x}_{\mathrm{A}},\boldsymbol{x}_{\mathrm{B}})$.
We denote the ChromaNet response for A by $\boldsymbol{y}_{\boldsymbol{\theta},{\mathrm{A}}} =
(g \circ f_{\boldsymbol{\theta}})(T_{c}\boldsymbol{x})$ idem for B.
The circular cross-correlation between $\boldsymbol{y}_{\boldsymbol{\theta},{\mathrm{A}}}$ and $\boldsymbol{y}_{\boldsymbol{\theta},{\mathrm{B}}}$ is
\begin{equation}
\boldsymbol{R}_{\boldsymbol{y}_{\boldsymbol{\theta},{\mathrm{A}}},\boldsymbol{y}_{\boldsymbol{\theta},{\mathrm{B}}}}[k] =
\sum_{q=0}^{Q-1} \boldsymbol{y}_{\boldsymbol{\theta},{\mathrm{A}}}[q] \boldsymbol{y}_{\boldsymbol{\theta},{\mathrm{B}}}[(q+k)\,\textrm{mod}\,Q]
\end{equation}
for $0\leq k<12$.
Taking the DFT of the equation above yields the circular cross-power spectral density (CPSD)
\begin{equation}
\widehat{\boldsymbol{R}}_{\boldsymbol{y}_{\boldsymbol{\theta},{\mathrm{A}}},\boldsymbol{y}_{\boldsymbol{\theta},{\mathrm{B}}}}[\omega] = \mathcal{F}\{\boldsymbol{R}_{\boldsymbol{y}_{\boldsymbol{\theta},{\mathrm{A}}},\boldsymbol{y}_{\boldsymbol{\theta},{\mathrm{B}}}}\}[\omega] = \widehat{\boldsymbol{y}}_{\boldsymbol{\theta},{\mathrm{A}}}[\omega]
\widehat{\boldsymbol{y}}_{\boldsymbol{\theta},{\mathrm{B}}}^{\ast}[\omega],
\end{equation}
where the asterisk denotes complex conjugation.

\subsection{Differentiable distance over the circle of fifths}
\label{sub:distance}

Given a constant DFT frequency $\omega=7$ and an arbitrary musical interval $k$ in semitones, we compute the CPSD associated to the pair $(\boldsymbol{y}_{\boldsymbol{\theta},\mathrm{A}},\boldsymbol{y}_{\boldsymbol{\theta},\mathrm{B}})$ and measure its half squared Euclidean distance to $e^{- 2\pi\mathrm{i}\omega k/Q}$ in the complex domain:
\begin{equation}
\mathcal{D}_{\boldsymbol{\theta},k}(\boldsymbol{x}_{\mathrm{A}}, \boldsymbol{x}_{\mathrm{B}}) = \dfrac{1}{2} \big \vert
e^{- 2\pi\mathrm{i}\omega k/Q} - \widehat{\boldsymbol{R}}_{ \boldsymbol{y}_{\boldsymbol{\theta},\mathrm{A}},\boldsymbol{y}_{\boldsymbol{\theta},\mathrm{B}}}[\omega]
\big \vert^2.
\label{eq:ddcf}
\end{equation}
Intuitively, in the case where $\widehat{\boldsymbol{y}}_{\boldsymbol{\theta},{\mathrm{A}}}[\omega]
$ and $\widehat{\boldsymbol{y}}_{\boldsymbol{\theta},{\mathrm{B}}}[\omega]$ are both one hot-encoding of 12 dimensions, they will be mapped as complex numbers of module 1 on the border of the CoF, $\widehat{\boldsymbol{R}}_{ \boldsymbol{y}_{\boldsymbol{\theta},\mathrm{A}},\boldsymbol{y}_{\boldsymbol{\theta},\mathrm{B}}}[\omega]$ measures the difference of phases on the CoF. Then, $\mathcal{D}_{\boldsymbol{\theta},k}(\boldsymbol{x}_{\mathrm{A}}, \boldsymbol{x}_{\mathrm{B}})$ measures its deviation from the DFT basis vector $e^{- 2\pi\mathrm{i}\omega k/Q}$, which corresponds to the actual pitch shift $k$ on the CoF. 
This distance is differentiable with respect to the weight vector $\boldsymbol{\theta}$.

\subsection{Invariance loss}
\label{section:losses}
Although the contents of $\boldsymbol{x}_{\mathrm{A}}$ versus $\boldsymbol{x}_{\mathrm{B}}$ may differ in terms of melody, rhythm, and instrumentation, we assume them to be in the same key.
This implies that ChromaNet responses $\boldsymbol{y}_{\boldsymbol{\theta},\mathrm{A}}$ and  $\boldsymbol{y}_{\boldsymbol{\theta},\mathrm{B}}$ should be maximally correlated at the unison interval $k=0$ and decorrelated for $k\neq0$.
In other words, the CPSD at the frequency $\omega$ should be maximal; i.e., equal to one.
Thus, given an arbitrary pitch interval $c$, we define an \emph{invariance loss} $\mathcal{L}_{\mathrm{AB}}$, defined as the distance between $T_c \boldsymbol{x}_{\mathrm{A}}$ and $T_c \boldsymbol{x}_{\mathrm{B}}$ on the CoF.
We obtain\footnote{In this paper, we use the vertical bar notation so as to clearly separate neural network parameters on the left versus data on the right.}:
\begin{align}
\mathcal{L}_{\mathrm{AB}}(\boldsymbol{\theta}\,\vert\,\boldsymbol{x},c) &= 
\mathcal{D}_{\boldsymbol{\theta},0}(T_c \boldsymbol{x}_{\mathrm{A}}, T_c \boldsymbol{x}_{\mathrm{B}}) 
\label{eq:invariance-loss}
\end{align}

\subsection{Equivariance loss}
\label{sub:equivariance-loss}
We want the model $f_{\boldsymbol{\theta}}$ to be equivariant to pitch transpositions.
Hence, we define an \emph{equivariance loss} $\mathcal{L}^{\mathrm{AA}}_{c,k}$ as the distance between $T_c \boldsymbol{x}_{\mathrm{A}}$ and $T_{(c+k)} \boldsymbol{x}_{\mathrm{A}}$ on the CoF:
\begin{align}
\mathcal{L}_{\mathrm{AA}}&(\boldsymbol{\theta}\,\vert\,\boldsymbol{x},c,k) = 
\mathcal{D}_{\boldsymbol{\theta},k}(T_c \boldsymbol{x}_{\mathrm{A}}, T_{c+k} \boldsymbol{x}_{\mathrm{A}}). 
\label{eq:equivariance-loss}
\end{align}
In theory, setting the architecture of $f_{\boldsymbol{\theta}}$ to a ChromaNet should lead to equivariance by design, for any value of the weight vector $\boldsymbol{\theta}$.
Yet, in practice, we observed that some values of $\boldsymbol{\theta}$ break this property of equivariance, likely due to boundary artifacts in 2-D convolutions---a similar observation to PESTO \cite{riou2023pesto}.
For STONE, only minimizing the invariance loss $\mathcal{L}^{\mathrm{AB}}_{c}$ causes the ChromaNet to collapse and predict a constant one-hot vector regardless of audio input $\boldsymbol{x}$ under certain hyperparameter choices, particularly for $\omega=1$.
To prevent this collapse, we penalize $f_{\boldsymbol{\theta}}$ with the equivariance loss in Equation \ref{eq:equivariance-loss}.

\subsection{Combined invariance and equivariance loss}
In addition, we penalize $f_{\boldsymbol{\theta}}$ according to the following loss:
\begin{align}
\mathcal{L}_{\mathrm{BA}}&(\boldsymbol{\theta}\,\vert\,\boldsymbol{x},c,k) = 
\mathcal{D}_{\boldsymbol{\theta},k}(T_c \boldsymbol{x}_{\mathrm{B}}, T_{c+k} \boldsymbol{x}_{\mathrm{A}}). 
\label{eq:combined-inv-eq-loss},
\end{align}
i.e., the distance between ChromaNet responses $T_{c} \boldsymbol{x}_{\mathrm{B}}$ and $T_{(c+k)} \boldsymbol{x}_{\mathrm{A}}$ on the CoF.
Observe that both these responses are already available after Equations \ref{eq:invariance-loss} and \ref{eq:equivariance-loss}.
Therefore, the inclusion of Equation \ref{eq:combined-inv-eq-loss} in the loss comes at almost no extra computational cost during gradient backpropagation.

\section{Self-supervised key signature profiles}
\label{sec:ssl}

\subsection{Training on real-world unlabeled data}
\label{sub:training}
We collect 60k songs from the catalog of a music streaming service, with due permission.
For each of them, we extract two disjoint segments $\boldsymbol{x}_{\mathbf{A}}$ and $\boldsymbol{x}_{\mathbf{B}}$ of duration equal to 15 seconds each.
Following prior knowledge in music cognition \cite{loui2007harmonic}, we set this duration to be as large as possible, considering the memory constraints of GPU hardware.

We implement ChromaNet and CPSD in PyTorch.
The interval $c$ (see Section \ref{sub:cqt}) varies between zero and 15 semitones while the interval $k$ (see Section \ref{sub:equivariance-loss}) varies between $-12$ and $12$ semitones.
We define a CPSD-based stochastic loss function by combining Equations \ref{eq:invariance-loss}, \ref{eq:equivariance-loss}, and \ref{eq:combined-inv-eq-loss}:
\begin{align}
\mathcal{L}^{\mathrm{CPSD}}(\boldsymbol{\theta}\,&\vert\,\boldsymbol{x},k,c) =
\mathcal{L}_{\mathrm{AB}}(\boldsymbol{\theta}\,\vert\,\boldsymbol{x},c)
\nonumber \\
&+
\mathcal{L}_{\mathrm{AA}}(\boldsymbol{\theta}\,\vert\,\boldsymbol{x},k,c) +
\mathcal{L}_{\mathrm{BA}}(\boldsymbol{\theta}\,\vert\,\boldsymbol{x},k,c),
\label{eq:cpsd-loss}
\end{align}
where CQT samples $\boldsymbol{x}$ and intervals $c$ and $k$ are drawn independently and uniformly at random.
We train the ChromaNet for 50 epochs using a cosine learning rate schedule with a linear warm-up.
We use an AdamW optimizer with a learning rate of $10^{-3}$ and a batch size of 128.

The self-supervised procedure above learns an approximately equivariant mapping from CQT to key signature profiles (KSP).
After training, we observe informally that for each input $\boldsymbol{x}$, most of the softmax activation in the KSP $\boldsymbol{y}_{\boldsymbol{\theta}}$ is concentrated on a single pitch class.
In other words, the loss $\mathcal{L}^{\mathrm{CPSD}}$ (Equation \ref{eq:cpsd-loss}) is sparsity-promoting.

\subsection{Calibration on a C major scale}
\label{sub:calibration}
By learning to predict pitch transpositions between segments, STONE learns the notion of relative tonality, just like the relative pitch of musicians; however, it lacks a notion of absolute tonality. 
Thanks to the equivariant property of ChromaNet, we only need to introduce this notion via a single recording of a C major scale paired with C major chords.
This calibration procedure resembles previous work in self-supervised fundamental frequency estimation\cite{gfeller2020spice, riou2023pesto}.

Denoting the $\texttt{C:maj}$ calibration sample by $\boldsymbol{x}_{\mathrm{cal}}$, we look up the index of its highest KSP coefficient in STONE:
\begin{equation}
q_{\mathrm{cal}}(\boldsymbol{\theta}) = \mathrm{arg}\max_{0\leq q' < Q} (g \circ f_{\boldsymbol{\theta}})(\boldsymbol{x}_{\mathrm{cal}})[q'].
\end{equation}
Then, given a CQT matrix $\boldsymbol{x}$ from the test set, we realign its ChromaNet response $\boldsymbol{y}=(g\circ f_{\boldsymbol{\theta}})(\boldsymbol{x})$ via a pitch transposition of the learned KSP by $q_{\mathrm{cal}}(\boldsymbol{\theta})$ semitones:
\begin{equation}
h_{\boldsymbol{\theta}}(\boldsymbol{y})[q] = \boldsymbol{y}[(q - q_{\mathrm{cal}}(\boldsymbol{\theta}))\,\mathrm{mod}\,Q].
\end{equation}

\subsection{Evaluation on real-world labeled data (FMAK)}
\label{sub:fmak}
FMAK is a subset of Free Music Archive dataset \cite{defferrard2017fma} containing 5489 songs that present a clear key and are in major or minor modes.
We label these songs by ear. Each of the 24 keys is represented in FMAK by at least 89 songs. $\texttt{C:maj}$ and $\texttt{A:min}$ are the best represented, while $\texttt{G\musSharp{}:maj}$ and $\texttt{G\musSharp{}:min}$ are the least represented. Songs are distributed in major and minor modes evenly. Rock and electronic dance music are the best represented genres; jazz and blues are the least represented.
To our knowledge, FMAK stands as the largest and most diverse MIR dataset with key annotation.

\subsection{Key signature estimation accuracy (KSEA)}
\label{sub:results-ksea}

In compliance with MIREX \cite{raffel2014mir_eval}, we propose the following figure of merit for key signature estimation:
\begin{align}
\mathrm{KSEA}(\boldsymbol{\theta}) &= \dfrac{1}{N} \sum_{n=0}^{N-1} \Big(
\boldsymbol{\delta}[s_{n}(\boldsymbol{\theta}) - S_{n}^{\mathrm{ref}}] \nonumber \\
 &+ \dfrac{1}{2} \boldsymbol{\delta}\big[\vert \vert s_{n}(\boldsymbol{\theta}) - S_{n}^{\mathrm{ref}} \vert - 6 \vert -1\big]\Big),
\end{align}
where $s_n(\boldsymbol{\theta}) = \mathrm{arg}\max_{0\leq q < Q}\big(h_{\boldsymbol{\theta}} \circ g \circ f_{\boldsymbol{\theta}}\big)(\boldsymbol{x}_{n})[q]
$ and $\boldsymbol{\delta}$ is the Kronecker symbol.
KSEA assigns a full point to the prediction if it matches the reference and a half point if the prediction is one perfect fifth above or below the reference.

\subsection{Results on key signature estimation}
We evaluate two variants of STONE on FMAK, all other things being equal: $\omega=7$ (CoF) and $\omega=1$.
For comparison, we also evaluate the work of Korzeniowski \textit{et al.}, the supervised state of of the art 
 (SOTA) for this task\cite{korzeniowski2018genre}, a convnet trained on GSMK.
Lastly, we evaluate a feature engineering pipeline, requiring no supervision: i.e., we take the global average of the chromagram representation and extract the pitch class with highest energy.

\begin{table}
    \centering
    \caption{Evaluation of self-supervised models on FMAK. KSEA denotes key signature estimation accuracy. We also report the supervised state of the art (SOTA) for comparison.}
    \begin{tabular}{l r r r}
         & Correct & Fifth & KSEA \\
        \hline
        Feature engineering & 1599 & 981 & 38\% \\
        STONE $(\omega=7)$& 3587 & 1225 & 77\%\\
        \bf{STONE} \boldsymbol{$(\omega=1)$}& \bf{3883} & 920 & \bf{79\%}\\
        \hline
        Supervised SOTA \cite{korzeniowski2018genre}& 4090 & 741 & 81\%
    \end{tabular}
    \label{tab:ssl-key-signature}
\end{table}

Table \ref{tab:ssl-key-signature} summarizes our results.
We observe that, for both values of the CPSD frequency $\omega$, STONE outperforms the feature engineering baseline.
Furthermore, for $\omega=1$, the KSEA approaches that of the supervised SOTA.

\subsection{Ablation study}
The two main novel components of STONE are the ChromaNet (Section \ref{sub:chromanet}) on one hand and cross-power spectral density (CPSD) over learned key signature profiles (KSP) on the other hand.
In order to evaluate their relative on performance, we conduct an ablation study: i.e., we substitute them by more conventional alternatives.

First, we replace the non-learned octave equivalence layer $g$ (Equation \ref{eq:softmax}) by a fully connected layer with same output size.
Secondly, we replace the three CPSD-based losses (Equation \ref{eq:cpsd-loss}) by $12$-class cross-entropy losses.
Intuitively, the first ablation disables equivariance in $f_{\boldsymbol{\theta}}$ while the second disables equivariance in $\mathcal{L}$.
We observe that both ablations cause a collapse of SSL, leading it to predict the majority class (i.e., $\texttt{C:maj}$) on almost every sample.
This suggests that both octave equivalence and CPSD are essential to the success of STONE.

\begin{table}
    \centering
    \caption{Ablation study of STONE ($\omega=7$) on FMAK. CPSD denotes cross-power spectral density. KSEA denotes key signature estimation accuracy. We report a naive baseline (i.e., predict the key signature of $\texttt{C:maj}$ and $\texttt{A:min}$ for every sample) for comparison.}
    \begin{tabular}{l r r r}
         & Correct & Fifth & KSEA \\
        \hline
        \textbf{STONE} $\boldsymbol{(\omega=7)}$ & \textbf{3587} & 1225 & \textbf{77\%}\\
        w/o octave equivalence & 1052 & 1267 & 31\%\\
        w/o CPSD & 1049  & 1267 & 31\%\\
        \hline
        Baseline (predict $\texttt{C}$) & 1049 & 1267 & 31\%\\
    \end{tabular}
    \label{tab:ablation-study}
\end{table}

\section{Self-supervised tonality estimation}
\label{sec:downstream}
\subsection{Structured prediction}
After having established that STONE learns to represent key signatures without any supervision, we turn to study its transferability to the well-known MIR problem of key estimation.
For this purpose, we must accommodate the distinction between major and minor keys, thus doubling the output dimension of the ChromaNet from 12 to 24.

We note that key signature and mode are orthogonal concepts: each major key has exactly one relative minor and vice versa.
These considerations suggest that the downstream task of key estimation may be formulated as structured prediction: i.e., in a 2-D label space.
We encode structured labels in a matrix $\mathbf{Y}$ with 12 rows and two columns.

\subsection{Batch normalization across key signatures}
\begin{figure}
    \centering
    \includegraphics[width=0.8\linewidth]{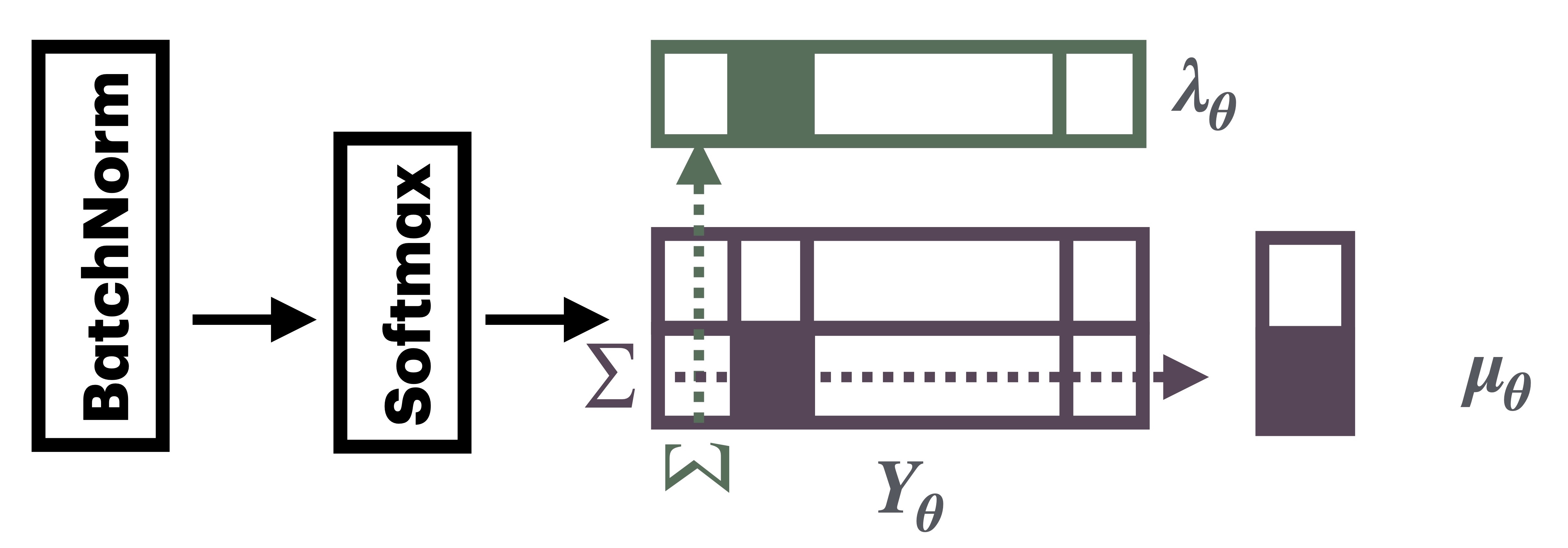}
    \caption{We modify the ChromaNet architecture of Figure \ref{fig:overview} to accommodate structured prediction key signature and mode.
    We apply batch normalization per mode $m$ and softmax over all coefficients, yielding a $12\times2$ matrix $\mathbf{Y}_{\boldsymbol{\theta}}(\boldsymbol{x})$. Summing $\mathbf{Y}_{\boldsymbol{\theta}}(\boldsymbol{x})$ over modes $m$ yields a learned key signature profile $\lambda_{\boldsymbol{\theta}}(\boldsymbol{x})$ in dimension 12; summing $\mathbf{Y}_{\boldsymbol{\theta}}(\boldsymbol{x})$ over chromas $q$ yields a pitch-invariant 2-dimensional vector $\mu_{\boldsymbol{\theta}}(\boldsymbol{x})$.}
    \label{fig:mode}
\end{figure}
We modify the last layer of the ChromaNet $f_{\boldsymbol{\theta}}$ to output two channels instead of one.
We also redefine the non-learnable operator $g$ for octave equivalence to accommodate two channels, apply batch normalization with non-learnable parameters on each channel, and a softmax nonlinearity over all batch-normalized coefficients.
This procedure normalizes each channel to null mean and unit variance over the training set, thus ensuring that both channels are activated and thus prevents a form of collapse during self-supervision. 

The composition of $g$ and $f_{\boldsymbol{\theta}}$, under their new definitions, yields a matrix $\mathbf{Y}_{\boldsymbol{\theta}}(\boldsymbol{x})$ with $Q=12$ rows and two columns.
By property of the softmax in $g$, all $24$ coefficients in  $\mathbf{Y}_{\boldsymbol{\theta}}(\boldsymbol{x})$ are positive and sum to one.
As illustrated in Figure \ref{fig:mode}, we take advantage of this property to derive a key signature estimator $\lambda_{\boldsymbol{\theta}}$ and a mode estimator $\mu_{\boldsymbol{\theta}}$, respectively defined as row-wise and column-wise partial sums of $\mathbf{Y}_{\boldsymbol{\theta}}(\boldsymbol{x})$:
\begin{equation}
  \lambda_{\boldsymbol{\theta}}(\boldsymbol{x})[q] = \sum_{m=0}^{1}\mathbf{Y}_{\boldsymbol{\theta}}(\boldsymbol{x})[q, m]
\end{equation}
\begin{equation}
  \mu_{\boldsymbol{\theta}}(\boldsymbol{x})[m] = \sum_{q=0}^{11}\mathbf{Y}_{\boldsymbol{\theta}}(\boldsymbol{x})[q, m].
  \label{eq:lambda-mu}
\end{equation}
We verify that the 12-dimensional vector $\lambda_{\boldsymbol{\theta}}(\boldsymbol{x})$ is positive, sums to one, and is \emph{equivariant} to pitch transpositions in $\mathbf{Y}_{\boldsymbol{\theta}}(\boldsymbol{x})$.
Conversely, the 2-dimensional vector $\mu_{\boldsymbol{\theta}}(\boldsymbol{x})$ is positive, sums to one, and is \emph{invariant} to pitch transpositions in $\mathbf{Y}_{\boldsymbol{\theta}}(\boldsymbol{x})$.
We use $\lambda_{\boldsymbol{\theta}}$ as a substitute for $(g\circ f_{\boldsymbol{\theta}})$ in $\mathcal{L}^{\mathrm{CPSD}}$.

\subsection{Self-supervised mode estimation}
\label{sub:bce}
We now introduce a loss for self-supervised mode estimation.
To this aim, we posit that mode is not only constant throughout the musical piece, but also remains invariant by pitch transposition.
Therefore, going back to the notations from Section \ref{sec:ssl}: $T_c \boldsymbol{x}_{\mathrm{A}}$, $T_c \boldsymbol{x}_{\mathrm{B}}$, and $T_{c+k} \boldsymbol{x}_{\mathrm{B}}$ should elicit the same value of the mode estimator $\mu_{\boldsymbol{\theta}}$.

We recall the definition of binary cross-entropy (BCE) for 2-D vectors whose entries are positive and sum to one:
\begin{equation}
\mathrm{BCE}(\boldsymbol{\mu}, \boldsymbol{\mu}') = - \boldsymbol{\mu}[0] \log \boldsymbol{\mu'}[0]
- \boldsymbol{\mu}[1] \log \boldsymbol{\mu'}[1].
\end{equation}
We compute the pairwise BCE between mode estimator responses a ssociated to the three predictions of the self-supervised ChromaNet (see Figure \ref{fig:overview})\footnote{Compared to $\mathcal{L}^{\mathrm{CPSD}}$,  we have swapped $\boldsymbol{x}_{\mathrm{A}}$ with $\boldsymbol{x}_{\mathrm{B}}$ in the first term. This is for compatibility with the supervised setting, as described in Section \ref{sec:semi-tones}, so as to avoid an undefined BCE due to a logarithm of zero.}:
\begin{align}
\mathcal{L}^{\mathrm{BCE}}(\boldsymbol{\theta}\,\vert\,\boldsymbol{x},c,k) &= \mathrm{BCE}(\mu_{\boldsymbol{\theta}}(T_{c}\boldsymbol{x}_{B}), \mu_{\boldsymbol{\theta}}(T_{c}\boldsymbol{x}_{A}))
\nonumber \\
&+ \mathrm{BCE}(\mu_{\boldsymbol{\theta}}(T_{c}\boldsymbol{x}_{A}), \mu_{\boldsymbol{\theta}}(T_{c+k}\boldsymbol{x}_{A}))
\nonumber \\
&+ \mathrm{BCE}(\mu_{\boldsymbol{\theta}}(T_{c}\boldsymbol{x}_{B}), \mu_{\boldsymbol{\theta}}(T_{c+k}\boldsymbol{x}_{A})).
\label{eq:bce-loss}
\end{align}

We add the BCE-based loss in Equation \ref{eq:bce-loss} to the CPSD-based loss in Equation \ref{eq:cpsd-loss}, thus yielding a full-fledged loss for self-supervised tonality estimation (STONE):
\begin{equation}
\mathcal{L}(\boldsymbol{\theta}\,\vert\,\boldsymbol{x},c,k) = \mathcal{L}^{\mathrm{CPSD}}(\boldsymbol{\theta}\,\vert\,\boldsymbol{x},c,k) + \mathcal{L}^{\mathrm{BCE}}(\boldsymbol{\theta}\,\vert\,\boldsymbol{x},c,k).
\end{equation}

We train the modified ChromaNet to minimize $\mathcal{L}$ with the same optimization hyperparameters as in Section \ref{sub:training}.
The resulting model, named 24-STONE, performs key signature estimation with $\lambda_{\boldsymbol{\theta}}$ and mode estimation with $\mu_{\boldsymbol{\theta}}$.
However these estimators are uncalibrated: i.e., $\lambda_{\boldsymbol{\theta}}$ only contains information of relative tonalities and $\mu_{\boldsymbol{\theta}}$ may swap relative major and minor.
We calibrate them by means of a $\texttt{C:maj}$ scale, via the same procedure as in Section \ref{sub:calibration}.


\subsection{Results on key and mode classification}
\label{sub:results-key-mode}
We evaluate two variants of 24-STONE on FMAK: $\omega=1$ and $\omega=7$; as well as the supervised SOTA.
We also evaluate the template matching algorithm of \cite{krumhansl2001cognitive}, requiring behavioral data but no supervision.

Table \ref{tab:ssl-key-mode} summarizes our results.
The 24-STONE model with $\omega=7$ is best in the unsupervised category, although well below the supervised SOTA.
However, setting $\omega$ to $1$ dramatically hurts the MIREX score of 24-STONE, placing it below the naive baseline.
Thus, formulating CPSD regression over the CoF (see Section \ref{sub:distance}) seems necessary for STONE to transfer to key and mode estimation, even so it is ouperformed by $\omega=1$ in KSEA (see Section  \ref{sub:results-ksea}).
With this result in mind, we set $\omega=7$ in the rest of this paper.

\begin{table*}
    \centering
    \begin{tabular}{l r r r r r r}
         & Correct & Fifth & Relative & Parallel & Wrong & MIREX \\ \hline
       Template matching \cite{krumhansl2001cognitive} & 2398 & 631 & 390 & 506 & 1564 & 53.4\% \\
       24-STONE ($\omega=1$) & 421 & 535 & 399 & 253 & 3881 & 15.6\%\\
        \textbf{24-STONE $\boldsymbol{(\omega=7)}$} & \textbf{2443} & 628 & 1320 & 115& 
    983 & \textbf{57.9\%} \\ \hline
       Supervised SOTA \cite{korzeniowski2018genre} & 3586 & 482 & 504 & 165 & 752 & 73.1\%\\
       Baseline (predict $\texttt{C:maj}$) & 551 & 568 & 498 & 286 & 3586 & 19.0\%\\
    \end{tabular}
    \caption{Evaluation of self-supervised models for key and mode estimation on FMAK. We also report the supervised state of the art (SOTA) \cite{korzeniowski2018genre} and a naive baseline (i.e., predict $\texttt{C:maj}$ for every sample) for comparison. See Section \ref{sub:results-key-mode} for details.}
    \label{tab:ssl-key-mode}
\end{table*}

\section{Semi-supervised tonality estimation}

\subsection{Supervising the ChromaNet}
\label{sec:semi-tones}
Thanks to structured prediction, the ChromaNet accommodates supervised training in the same label space as self-supervised training.
Note that the 24-STONE losses $\mathcal{L}^{\mathrm{CPSD}}$ and $\mathcal{L}^{\mathrm{BCE}}$ involves pairwise comparisons between three items belonging to the same $\boldsymbol{x}$: i.e., two transposed versions of the same segment $\boldsymbol{x}_{\mathrm{A}}$ and one from another segment $\boldsymbol{x}_{\mathrm{B}}$.
In this context, a simple way to introduce supervision is to replace the responses $\lambda_{\boldsymbol{\theta},\mathrm{B}}$ and $\mu_{\boldsymbol{\theta},\mathrm{B}}$ by ``oracles'' $\lambda_{\mathrm{ref}}$ and $\mu_{\mathrm{ref}}$ which are informed by the ground truth.

Given the ground truth key signature $q_{\mathrm{ref}}$ and mode $m_{\mathrm{ref}}$, one-hot encoding yields the sparse vectors
$\lambda_{\mathrm{ref},c}(\boldsymbol{x})[q] = \boldsymbol{\delta}[(q-q_{\mathrm{ref}}-c)\textrm{ mod 12}]$ and $
\mu_{\mathrm{ref}}(\boldsymbol{x})[m] = \boldsymbol{\delta}[m-m_{\mathrm{ref}}]$, where $c$ is a pitch interval in semitones (see Section \ref{sub:cqt}).
We use these oracles to write supervised variants of losses $\mathcal{L}^{\mathrm{CPSD}}$ and $\mathcal{L}^{\mathrm{BCE}}$.
This seamless switch from SSL to supervised learning requires no change of architecture nor optimizer.
Thus, instead of using supervision as fine-tuning, we propose an alternated scheme: one epoch of SSL followed by one epoch of supervised learning, and so on.

\subsection{Semi-TONE and Sup-TONE}
Introducing supervision into 24-STONE yields a semi-supervised tonality estimator, or Semi-TONE for short.
We alternate between self-supervised epochs on 60k unlabeled recordings (see Section \ref{sub:training}) and supervised epochs on the 1159 songs in  GSMK in which annotators agree.
For the sake of comparison, we experiment with disabling SSL and training the ChromaNet directly on GSMK: hence a fully supervised tonality estimator, or Sup-TONE for short.

To compare their ability to learn from limited labeled data, we retrain Semi-TONE and Sup-TONE after subsampling GSMK at random by factors of $10$ and $100$.

\begin{figure}[h]
\centering{\includegraphics[width=0.85\linewidth]{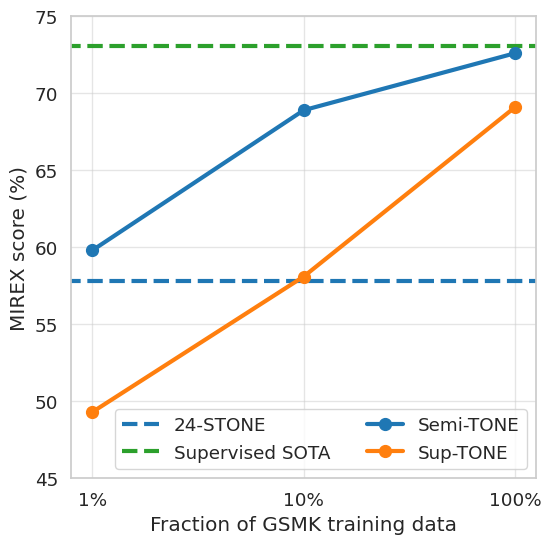}}
\caption{Evaluation of self-supervised (dashed blue), semi-supervised (solid blue), and supervised models (orange) on FMAK. All models use $\omega=7$. We also report the supervised state of the art (SOTA) \cite{korzeniowski2018genre} in dashed green.}
    \label{fig:mirex}
\end{figure}

\subsection{Results on key and mode classification}
Figure \ref{fig:mirex} summarizes our results.
We observe that Semi-TONE systematically outperforms Sup-TONE at any amount of training data.
In particular, training Semi-TONE with 10\% of GSMK leads to a comparable MIREX score as training Sup-TONE with 100\% of GSMK.
This result confirms the interest of our proposed pretext tasks towards the overarching goal of reducing human annotation effort.

\begin{figure}
    \centering
    \includegraphics[width=\linewidth]{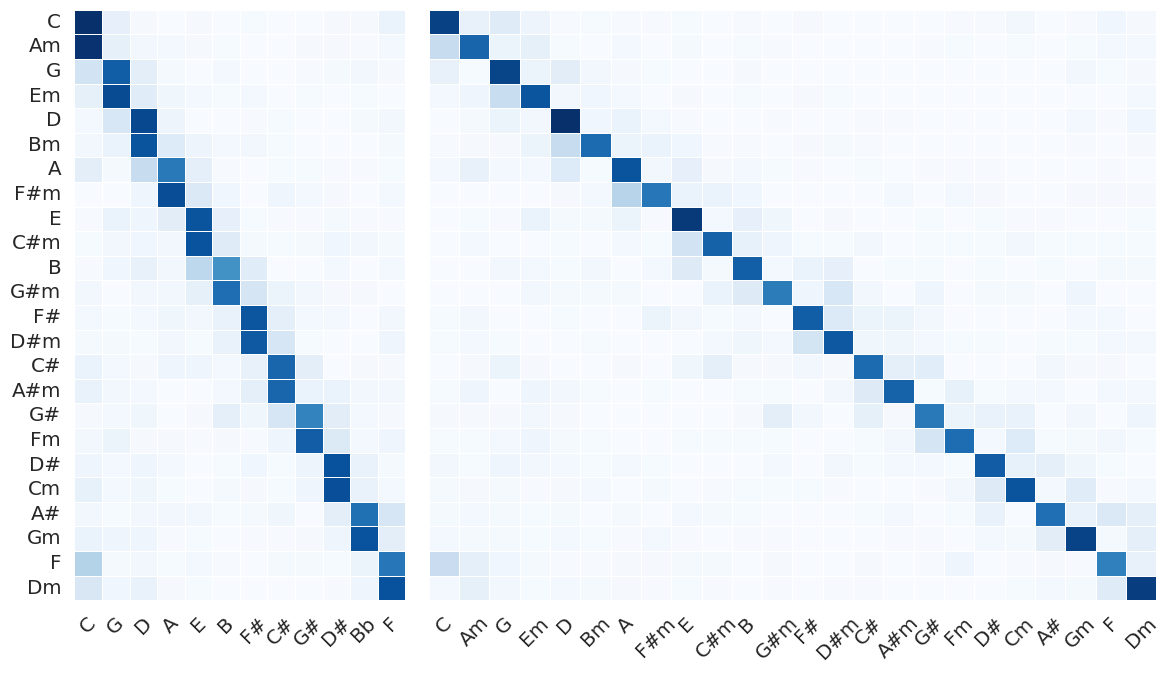}
    \caption{Confusion matrices of STONE (left, 12 classes) and Semi-TONE (right, 24 classes) on FMAK, both using $\omega=7$. The axis correspond to model prediction and reference respectively, keys arranged by proximity in the CoF and relative modes. Deeper colors indicate more frequent occurences per relative occurence per reference key.}
    \label{fig:confusion-matrix}
\end{figure}

Training Semi-TONE on the full GSMK dataset yields a MIREX score of 72.6\%; i.e., roughly on par with the supervised SOTA (73.1\%).
Figure \ref{fig:confusion-matrix} shows the confusion matrix of calibrated STONE and Semi-TONE on FMAK.
Although our methods do not outperform the SOTA on key estimation, it brings insights into a novel framework that does not require high supervision for training.
Moreover, we note that self-supervision remains beneficial even when the full GSMK dataset is available for training.
Therefore, a promising avenue of research is to scale up the dataset of unlabeled recordings (see Section \ref{sub:training}), thus widening the gap between Semi-TONE and Sup-TONE on FMAK.

\section{Conclusion}
STONE learns key signature profiles (KSP) via equivariant self-supervised learning in the time--frequency domain.
We have seen than a semi-supervised extension of STONE (semi-TONE) reduces expert annotation by 90\% less at no loss of MIREX score compared to the fully supervised variant (sup-TONE). 
The primary limitation of our work resides in the inability of the STONE objective (CPSD, i.e., cross-power spectral density) to distinguish major keys from minor keys.
Future work will study how STONE can be adapted to other pitch-relative MIR tasks. 

\bibliography{ISMIRtemplate}

\begin{thebibliography}{10}
\providecommand{\url}[1]{#1}
\csname url@samestyle\endcsname
\providecommand{\newblock}{\relax}
\providecommand{\bibinfo}[2]{#2}
\providecommand{\BIBentrySTDinterwordspacing}{\spaceskip=0pt\relax}
\providecommand{\BIBentryALTinterwordstretchfactor}{4}
\providecommand{\BIBentryALTinterwordspacing}{\spaceskip=\fontdimen2\font plus
\BIBentryALTinterwordstretchfactor\fontdimen3\font minus \fontdimen4\font\relax}
\providecommand{\BIBforeignlanguage}[2]{{%
\expandafter\ifx\csname l@#1\endcsname\relax
\typeout{** WARNING: IEEEtran.bst: No hyphenation pattern has been}%
\typeout{** loaded for the language `#1'. Using the pattern for}%
\typeout{** the default language instead.}%
\else
\language=\csname l@#1\endcsname
\fi
#2}}
\providecommand{\BIBdecl}{\relax}
\BIBdecl

\bibitem{mullensiefen2014musicality}
D.~M{\"u}llensiefen, B.~Gingras, J.~Musil, and L.~Stewart, ``{The musicality of non-musicians: An index for assessing musical sophistication in the general population},'' \emph{PLOS ONE}, vol.~9, no.~2, p. e89642, 2014.

\bibitem{zhu2021musicbert}
H.~Zhu, Y.~Niu, D.~Fu, and H.~Wang, ``{MusicBERT}: A self-supervised learning of music representation,'' in \emph{Proceedings of the ACM International Conference on Multimedia (MM)}, 2021, pp. 3955--3963.

\bibitem{spijkervet2021contrastive}
J.~Spijkervet and J.~A. Burgoyne, ``Contrastive learning of musical representations,'' in \emph{Proceedings of the International Society for Music Information Retrieval Late-Breaking/Demo Session (ISMIR-LBD)}, 2021.

\bibitem{desblancs2023zero}
D.~Desblancs, V.~Lostanlen, and R.~Hennequin, ``Zero-note samba: Self-supervised beat tracking,'' \emph{IEEE/ACM Transactions on Audio, Speech, and Language Processing}, 2023.

\bibitem{meseguer2024}
G.~Meseguer-Brocal, D.~Desblancs, and R.~Hennequin, ``An experimental comparison of multi-view self-supervised methods for music tagging,'' in \emph{Proceedings of the IEEE International Conference on Audio, Speech and Signal Processing (ICASSP)}, 2024.

\bibitem{schneider2019wav2vec}
S.~Schneider, A.~Baevski, R.~Collobert, and M.~Auli, ``{wav2vec: Unsupervised Pre-training for Speech Recognition},'' in \emph{Proceedings of INTERSPEECH}, 2019.

\bibitem{saeed2021contrastive}
A.~Saeed, D.~Grangier, and N.~Zeghidour, ``Contrastive learning of general-purpose audio representations,'' in \emph{Proceedings of the IEEE International Conference on Acoustics, Speech and Signal Processing (ICASSP)}, 2021.

\bibitem{niizumi2021byol}
D.~Niizumi, D.~Takeuchi, Y.~Ohishi, N.~Harada, and K.~Kashino, ``{BYOL} for audio: Self-supervised learning for general-purpose audio representation,'' in \emph{Proceedings of the IEEE International Joint Conference on Neural Networks (IJCNN)}, 2021.

\bibitem{riou2023pesto}
A.~Riou, S.~Lattner, G.~Hadjeres, and G.~Peeters, ``{PESTO}: Pitch estimation with self-supervised transposition-equivariant objective,'' in \emph{Proceedings from the International Society for Music Information Retrieval Conference (ISMIR)}, 2023.

\bibitem{quinton2022equivariant}
E.~Quinton, ``Equivariant self-supervision for musical tempo estimation,'' in \emph{Proceedings of the International Society for Music Information Retrieval Conference (ISMIR)}, 2022.

\bibitem{wong2023fmak}
S.~Wong and G.~Hernandez, ``Fmak: A dataset of key and mode annotations for the free music archive--extended abstract,'' in \emph{Proc. of the International Society for Music Information Retrieval Late-Breaking/Demo Session (ISMIR-LBD)}, 2023.

\bibitem{gfeller2020spice}
B.~Gfeller, C.~Frank, D.~Roblek, M.~Sharifi, M.~Tagliasacchi, and M.~Velimirovi{\'c}, ``{SPICE}: Self-supervised pitch estimation,'' \emph{IEEE/ACM Transactions on Audio, Speech, and Language Processing}, vol.~28, pp. 1118--1128, 2020.

\bibitem{morais2023tempo}
G.~Morais, M.~E. Davies, M.~Queiroz, and M.~Fuentes, ``Tempo vs. pitch: understanding self-supervised tempo estimation,'' in \emph{Proceedings of the IEEE International Conference on Acoustics, Speech and Signal Processing (ICASSP)}, 2023, pp. 1--5.

\bibitem{cwitkowitz2024toward}
F.~Cwitkowitz and Z.~Duan, ``Toward fully self-supervised multi-pitch estimation,'' \emph{arXiv preprint arXiv:2402.15569}, 2024.

\bibitem{noland2007signal}
K.~Noland and M.~Sandler, ``Signal processing parameters for tonality estimation,'' in \emph{Proceedings of the Audio Engineering Society Convention (AES)}, 2007.

\bibitem{pauws2004musical}
S.~Pauws, ``Musical key extraction from audio,'' in \emph{Proceedings of the International Society for Music Information Retrieval Conference (ISMIR)}, 2004.

\bibitem{faraldo2016key}
{\'A}.~Faraldo, E.~G{\'o}mez, S.~Jord{\`a}, and P.~Herrera, ``Key estimation in electronic dance music,'' in \emph{Proceedings of the European Conference on Information Retrieval (ECIR)}, 2016.

\bibitem{krumhansl2001cognitive}
C.~L. Krumhansl, \emph{Cognitive foundations of musical pitch}.\hskip 1em plus 0.5em minus 0.4em\relax Oxford University Press, 2001.

\bibitem{Wu2020AVA}
Y.~Wu, E.~Nakamura, and K.~Yoshii, ``A variational autoencoder for joint chord and key estimation from audio chromagrams,'' in \emph{Proceedings of the Asia-Pacific Signal and Information Processing Association Annual Summit and Conference (APSIPA ASC)}, 2020.

\bibitem{Wu2022JointCA}
Y.~Wu and K.~Yoshii, ``Joint chord and key estimation based on a hierarchical variational autoencoder with multi-task learning,'' \emph{APSIPA Transactions on Signal and Information Processing}, 2022.

\bibitem{Schreiber2019MusicalTA}
H.~Schreiber and M.~M{\"u}ller, ``Musical tempo and key estimation using convolutional neural networks with directional filters,'' in \emph{Proceedings of the International Sound and Music Computing Conference (SMC)}, 2019.

\bibitem{korzeniowski2018genre}
F.~Korzeniowski and G.~Widmer, ``Genre-agnostic key classification with convolutional neural networks,'' in \emph{Proceedings of the International Society on Music Information Conference (ISMIR)}, 2018.

\bibitem{Bertin-Mahieux2011}
T.~Bertin-Mahieux, D.~P. Ellis, B.~Whitman, and P.~Lamere, ``The million song dataset,'' in \emph{Proceedings of the International Society on Music Information Retrieval Conference ({ISMIR})}, 2011.

\bibitem{Knees2015TwoDS}
\BIBentryALTinterwordspacing
P.~Knees, {\'A}.~Faraldo, P.~Herrera, R.~Vogl, S.~B{\"o}ck, F.~H{\"o}rschl{\"a}ger, and M.~L. Goff, ``Two data sets for tempo estimation and key detection in electronic dance music annotated from user corrections,'' in \emph{International Society for Music Information Retrieval Conference}, 2015. [Online]. Available: \url{https://api.semanticscholar.org/CorpusID:15836728}
\BIBentrySTDinterwordspacing

\bibitem{burgoyne-2011-expert}
J.~A. Burgoyne, J.~Wild, and I.~Fujinaga, ``An expert ground truth set for audio chord recognition and music analysis,'' in \emph{Proceedings of the International Society for Music Information Retrieval Conference (ISMIR)}, A.~Klapuri and C.~Leider, Eds.\hskip 1em plus 0.5em minus 0.4em\relax University of Miami, 2011.

\bibitem{liu2022convnet}
Z.~Liu, H.~Mao, C.-Y. Wu, C.~Feichtenhofer, T.~Darrell, and S.~Xie, ``{A ConvNet for the 2020s},'' in \emph{Proceedings of the IEEE/CVF conference on computer vision and pattern recognition (CVPR)}, 2022, pp. 11\,976--11\,986.

\bibitem{muller2011signal}
M.~M{\"u}ller, D.~P. Ellis, A.~Klapuri, and G.~Richard, ``Signal processing for music analysis,'' \emph{IEEE Journal of Selected Topics in Signal Processing}, vol.~5, no.~6, pp. 1088--1110, 2011.

\bibitem{loui2007harmonic}
P.~Loui and D.~Wessel, ``{Harmonic expectation and affect in Western music: Effects of attention and training},'' \emph{Perception \& Psychophysics}, vol.~69, pp. 1084--1092, 2007.

\bibitem{defferrard2017fma}
M.~Defferrard, K.~Benzi, P.~Vandergheynst, and X.~Bresson, ``{FMA}: A dataset for music analysis,'' in \emph{Proceedings of the International Society for Music Information Retrieval Conference (ISMIR)}, 2017.

\bibitem{raffel2014mir_eval}
C.~Raffel, B.~McFee, E.~J. Humphrey, J.~Salamon, O.~Nieto, D.~Liang, D.~P. Ellis, and C.~C. Raffel, ``{mir\_eval: A Transparent Implementation of Common MIR Metrics.}'' in \emph{Proceedings of the International Society for Music Information Retrieval Conference (ISMIR)}, 2014.

\end{thebibliography}

\end{document}